# Classification of COVID-19 on chest X-Ray images using Deep Learning model with Histogram Equalization and Lungs Segmentation

Aman Swaraj[1*], Karan Verma[2]

aman_s@cs.iitr.ac.in[1*], karanverma@nitdelhi.ac.in[2]

Indian Institute of Technology Roorkee[1]; National Institute of Technology Delhi[2]

**\*Corresponding Author:**

Aman Swaraj
Ph.D. Research Scholar,
Indian Institute of Technology, Roorkee

Institute Address:

Indian Institute of Technology Roorkee
Roorkee, Uttarakhand
India - 247667

Permanent Address:
488- Solanipuram,
Roorkee- 247667
Uttarakhand, India

Tel: 91-7217842795, 91-7783860847
Email: aman_s@cs.iitr.ac.in ; amanswaraj007@gmail.com

**Total number of words of the manuscript- 6577**

**The number of words of the abstract- 238**

**The number of figures- 8**

**The number of tables- 10**

**ABSTRACT**

**Background and Objective**: Artificial intelligence (AI) methods coupled with biomedical analysis has a critical role during pandemics as it helps to release the overwhelming pressure from healthcare systems and physicians. Radiological imaging can act as an important diagnostic tool to accurately classify covid-19 patients and prescribe the necessary treatment in due time. Lung Segmentation plays a major role in getting accurate results as it prevents models learn features from outside the infected area. However, most of the existing work has not considered segmentation, and on the contrary have used impractical and dis-proportionate augmentation techniques. A more nuanced study with a balanced dataset without using any augmentation technique and segmentation of lungs is needed. With this motivation, we present a methodology to predict covid-19 in lungs X-rays more accurately and reliably.

**Dataset:** At the time of research, there was availability of around 470 covid-19 images. Our work compromises of a total of 2470 images for three different class labels, namely, healthy lungs, ordinary pneumonia, and covid-19 infected pneumonia, out of which 470 X-ray images belong to the covid-19 category.

**Methods:** We first pre-process all the images using histogram equalization techniques and segment them using U-net architecture. VGG-16 network is then used for feature extraction from the pre-processed images which is further sampled by SMOTE oversampling technique to achieve a balanced dataset. Finally, the class-balanced features are classified using a support vector machine (SVM) classifier with 10-fold cross-validation and the accuracy is evaluated.

**Result and Conclusion:** Our novel approach combining well-known pre-processing techniques, feature extraction methods, and dataset balancing method, lead us to an outstanding rate of recognition of 98% for COVID-19 images over a dataset of 2470 X-ray images. Our model achieves significant performance with respect to other state of art works.

## 1. INTRODUCTION

COVID-19 first appeared in late 2019 after an upsurge of peculiar pneumonia cases in Wuhan, China. With very few parallels in history, the covid-19 outbreak has already infected over 96 million people and caused more than 2 million deaths worldwide [1]. Owing to its similarities with previous outbreaks such as SARS and MERS [2, 3], the covid-19 virus has been linked with the possibility of having a zoonotic nature [4, 5]. Characterized by the WHO as a global pandemic [6] the virus continues to spread with a brisk rate of around 1.4 to 2.5 [7].

To slow down the outbreak, the nationwide lockdown has been observed in almost every country throughout the globe. However, in densely populated countries like India, Brazil, and the United States, the curve has continued to jump even after employing strict intervention policies. Therefore, rapid clinical tools for detecting the coronavirus have become the preeminent need for most healthcare facilities now. In our earlier work, we developed a hybrid model to forecast future covid cases [8, 9]. Although, a forecasting model can help slow down the spread, it can't diagnose patients in particular.

The most common technique being currently adopted for diagnosing covid-19 patients is known as real-time reverse transcription-polymerase chain reaction (RT-PCR). However, RT-PCR is reported to have an average sensitivity around 60% only, and therefore to ensure the complete diagnosis of the potential patient, chest X-rays have been equally recommended important for covid-19 detection [10-14].

In recent times, artificial intelligence (AI) has emerged as the foremost catalyst in advancing biomedical research [15,16]. Deep learning approaches have been widely used in applications involving medical image analysis such as skin cancer classification [17], breast cancer detection [18], Electroencephalography (EEG) based diagnosis [19-22], brain diseases [23-25] etc. Since covid-19 also involves screening of chest X-rays, deep learning-based diagnosis of the lungs can aid the radiologists with quick and precise detection of the symptoms in a potential patient.

In this connection, Zhang et al. [26] proposed a confidence-aware anomaly detection (CAAD) model equipped with an anomaly detection module, a shared feature extractor, and a module for confidence prediction. Kumar et al. [27] also proposed a machine learning-based classification of extracted features of chest X-ray images using ResNet152. Wong et al. [28] obtained 92.4% accuracy with their tailor-made model 'COVID-Net'. In [29], Ozturk et al. proposed a model having an end-to-end architecture for binary as well as multi-class classification achieving 98% and 87% accuracy respectively. Toğaçar et al. [30] used fuzzy and stacking preprocessing techniques on the images before training them with MobileNetV2 and SqueezeNet models. Ucar et al. [31] also implemented the SqueezeNet model with Bayesian optimization on a dataset having 76 covid infected lung images. Similarly, several other works also include deep learning models employed on X-ray images for covid-19 classification [32-42]. We further draw a comparison between these models in our discussion section.

While all these studies have their significant contributions, these models are subject to anomalies as most of these models have no specific mechanism for segregating the lungs from the original X-ray images before extracting features from them, and therefore features detected from outside the lungs portion may result in misclassification. Adding to this, most of these models have been implemented on either a very scarce dataset or augmented dataset subject to overfitting. Hence, there is a need for a much elaborate study on this subject with appropriate models, pre-processing techniques, and similar supplementary methods for achieving more reliable results.

In this study, we run our model on a total of 470 X-ray images of COVID-19 infected lungs along with 1000 images each of ordinary pneumonia and healthy chest X-rays. For image enhancement, we first applied histogram equalization on all the images and then used U-Net [43] to isolate the lungs portion from the whole image. After preprocessing, we trained a modified VGG-16 model from scratch on 60% of our data and then to deploy the model as an encoder, the Soft-max layer of the trained model was removed. All the processed input images were then passed through the encoder to obtain 1024-dimensional feature vectors. Further, we utilize the SMOTE technique [44] on these feature vectors to obtain a proper class balanced dataset. Finally, the balanced dataset was classified using an SVM classifier with RBF kernel before evaluating the model with 10-fold validation.

The rest of the paper is structured as follows: In section 2, we elaborate on all the methods carried out in the study along with the description of the dataset. In section 3, we present the experiment analysis and results. Section 4 holds a discussion and section 5 depicts the conclusion.

## 2. METHODOLOGY

A brief overview of our proposed framework is depicted in Fig. 1. It can be broadly classified into the following four stages i.e., dataset collection, image pre-processing, feature extraction, and classification.

Section 2.1 provides a detailed description of the data sources. In section 2.2, image pre-processing steps are described and section 2.3 talks about feature extraction. Finally, in section 2.4, classification and evaluation techniques are discussed.

### 2.1 Dataset Description

For this study, we have used the publicly available dataset COVIDx [28] which at the time comprised of a total of 15371 posterior-to-anterior chest x-ray images. All these images are a combination of an improved version of several existing datasets. The class distribution of the original COVIDx dataset is tabulated in Table 1. COVID-19 is a recent disease, therefore, the images corresponding to its class are relatively low in number and may cause overfitting in prediction models. To handle this imbalance, we randomly selected 1000 images each of normal lungs and infected with ordinary pneumonia from the dataset. The final class distribution for the sampled dataset used in our study is mentioned in Table 2.

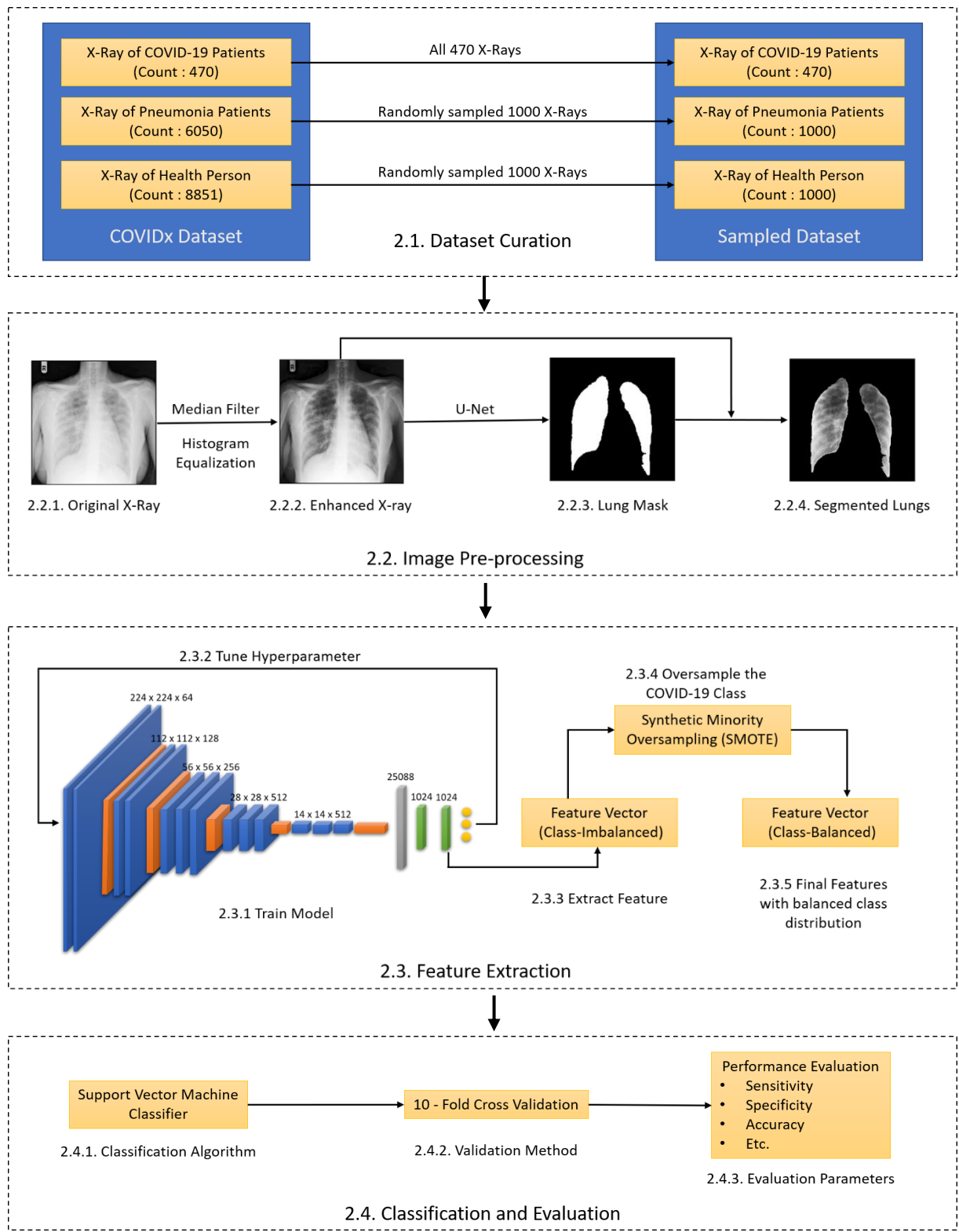

*Fig.1: Block diagram depicting our proposed framework.*

Data augmentation was avoided to prevent overfitting. Considering the large imbalance in data, using data augmentation to generate 600 images from the existing 400 images of the COVID-19 class could result in a loss of generalization. Consequently, to handle class imbalance, we have used class-weights. Class weights were determined empirically by an intuitive approach that loss corresponding to misclassification of COVID-19 image must be highly penalized.

*Table 1: Class Distribution of COVIDx dataset*

| Class | Images | Patient |
| --- | --- | --- |
| Normal | 8851 | 8851 |
| Pneumonia | 6050 | 6033 |
| COVID-19 | 470 | 330 |
| Total | 15371 | 15214 |

*Table 2: Class Distribution of sampled COVIDx dataset used in study*

| Class | Images | Patient |
| --- | --- | --- |
| Normal | 1000 | 1000 |
| Pneumonia | 1000 | 1000 |
| COVID-19 | 470 | 330 |
| Total | 2470 | 2330 |

## 2.2 Image pre-processing

All the X-ray images are first turned into 224x224 dimension images. Then, to enhance the images, the histogram equalization method is applied over a median filter. The enhanced images are then segmented using a U-Net convolutional neural network to obtain lung masks. The obtained lung mask is then used to segregate the lungs and thus the final dataset is prepared. The entire pre-processing flow is shown in Block-2.2 of Fig. 1.

### 2.2.1 Image Enhancement

It can be observed that there is high variability in the quality of the images. The dataset contains images of varying resolutions, shapes, and sizes. Moreover, some of the images are extremely low in quality. To effectively perform the classification of images, proper pre-processing techniques must be employed. And so, all the images were brought to scale at a size of 224 x 224. Further, a median filter was applied to make the images noise free.

To improve the quality of images we experimented with the various image enhancement techniques such as Laplacian Unsharp Filter, Gaussian Unsharp Filter, Histogram Equalization (HE) and Contrast Limited Adaptive Histogram Equalization (CLAHE). Each of these techniques were applied to three images from the dataset of varying quality (clarity). The results of these techniques are shown in Fig 2.

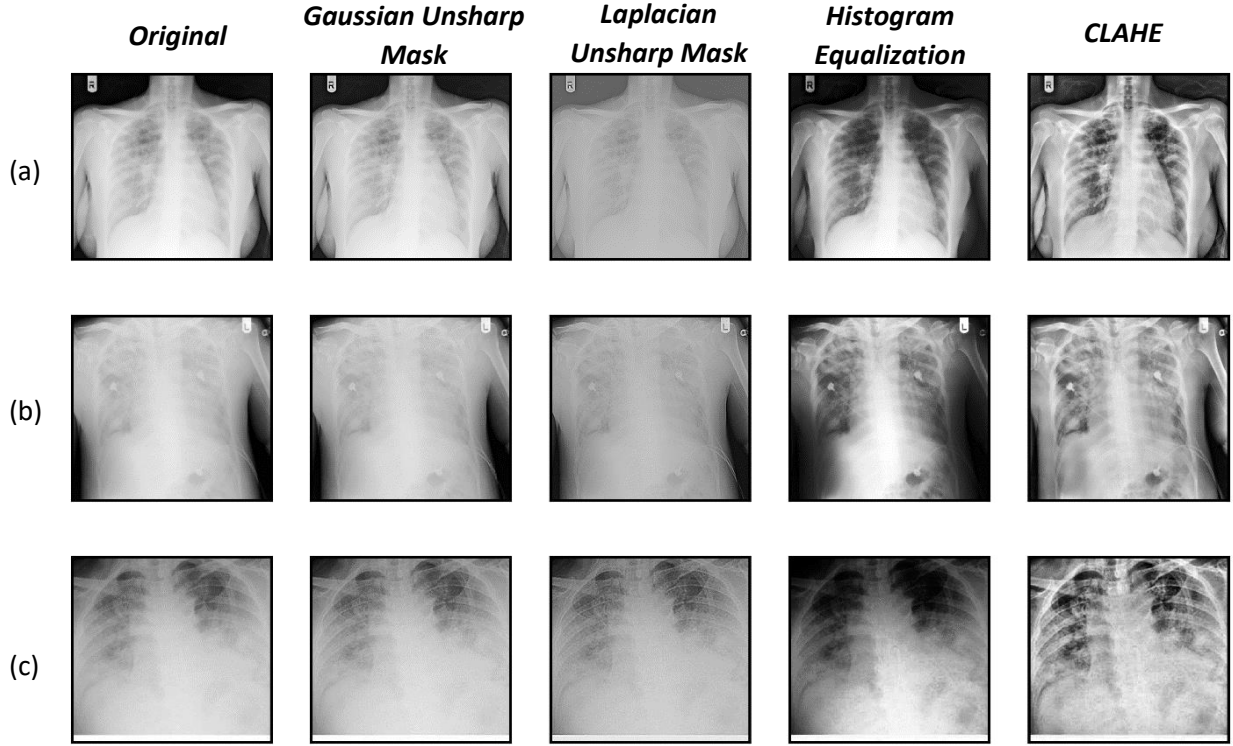

*Fig.2: Various image enhancement techniques.*

The primary motive of image enhancement at this stage is to facilitate the segmentation of the lung. It can be observed from Fig 3 that HE and CLAHE have shown remarkable results in denoising and enhancing the images. However, it can also be observed that applying CLAHE increases the intensity of bones as well which might later affect the performance of the lung segmentation and classification model since the neural network might recognize these (ribs and sternum bones) as the primary feature in the detection of COVID-19.

Hence, Histogram Equalization was selected as the image enhancement technique in our proposed framework.

### 2.2.2 Lung Segmentation

Gianluca Maguolo et al. in their study [45], pointed out a significant challenge while identifying COVID-19 infection using chest X-ray images. In their setup, they removed the lungs portion of the X-ray images by turning the corresponding pixels associated with the lungs completely black and then training their classifier on the resulting images. However, even after removing the lungs in this manner, they got similar results as with the lungs present. This is attributed to the fact that the neural networks can learn patterns that are not correlated to the infection region of the lungs in particular.

We present two primary arguments to justify the need for lung segmentation-

- Firstly, the neural networks might detect irrelevant features and still produce significantly high accuracy in COVID-19 detection. For instance, consider the X-rays of patients who are covid positive as displayed in Fig 3. It can be observed that the X-rays contain some extra features like the presence of ECG leads, wires, pacemakers, etc. A lot of the X-ray images have ECG leads on their X-ray since the patients were kept on ventilation given their critical situation. On an estimate, 30% of the images belonging to the COVID-19 criteria contain these features. Comparatively, the images of other 2 classes contain very few of these unnecessary features thus making the model highly unreliable.

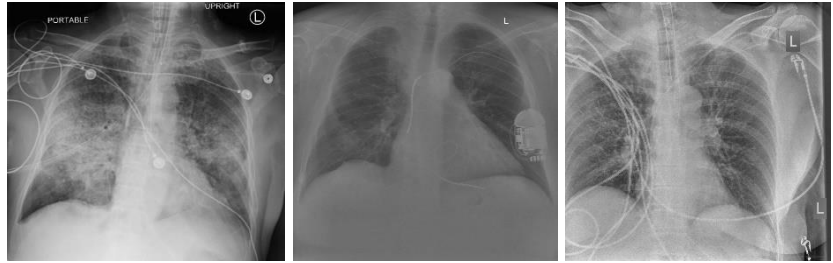

*Fig 3: Presence of ECG wires, leads, pacemakers.*

- Secondly, the number of X-ray images available for COVID-19 patients are relatively scarce. Also, as pointed out by authors of [45], the images available for the COVID-19 class are uploaded by doctors across the globe (multiple sources) while the images for the Normal and Pneumonia class belong to curated datasets (single source).

  For example, the dataset created by Dr. Paul Mooney for a Kaggle competition [46] on bacterial and viral pneumonia classification consists of pediatric X-ray images. The images from the dataset are naturally different from that of other images belonging to the COVID-19 class (the former belongs to children while the latter belongs primarily to adults).

  Therefore, the studies using images from this dataset along with COVID-19 class images might get significantly high accuracy not because of the distinction due to globular opacity in the lungs due to COVID-19 but because of natural distinction in the source of data itself.

However, eliminating dataset-dependent features can reduce inter-dataset differences. And, one of the ways this can be achieved is to use lung-segmentation. In our work, we implemented U-Net [47] which is a fully convolutional neural network usually employed for semantic segmentation in biomedical images. The fact that it requires fewer training samples and offers precise segmentation gives it an edge over other segmentation methods. Resultant images after lung segmentation are presented in Fig 4.

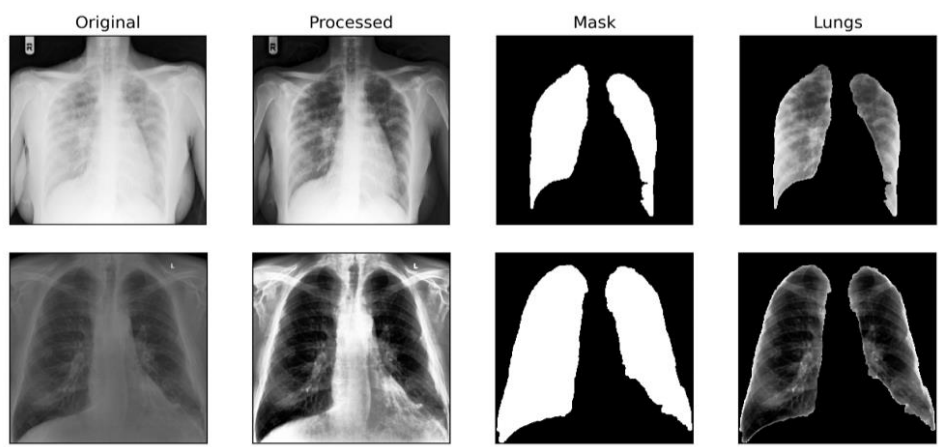

*Fig 4: Lung Segmentation Result (Original X-ray, Enhanced X-ray, Lung segmentation Mask, Lungs).*

## 2.3 Feature Extraction

Feature extraction in a medical dataset is a sensitive task and should be done with proper care. Since a radiograph contains both low-level features like texture, edges, blobs, shape, color, and high-level features like the thoracic cage, mediastinum, ECG wires, etc., selection of a proper model is necessary to segregate desired features.

Kong et.al, in their study [48], presented certain evidence that points to a covid-19 infection in a patient. Their study included evidence of bilateral nodular, peripheral ground-glass opacities, and consolidators as the prime indicators of covid-19. In radiologic terms, 'ground-glass' refers to hazy lung opacity having a thin appearance on the X-ray image, unable to obscure any underlying bronchial walls or pulmonary vessels. However, consolidation is the exact opposite of having dense opacities obscuring bronchial walls and vessels. The appearance of these indications in the chest radiograph of a COVID-19 patient is shown in Fig 5 which is taken from a study conducted by Rabab Yasin and Walaa Gouda [49].

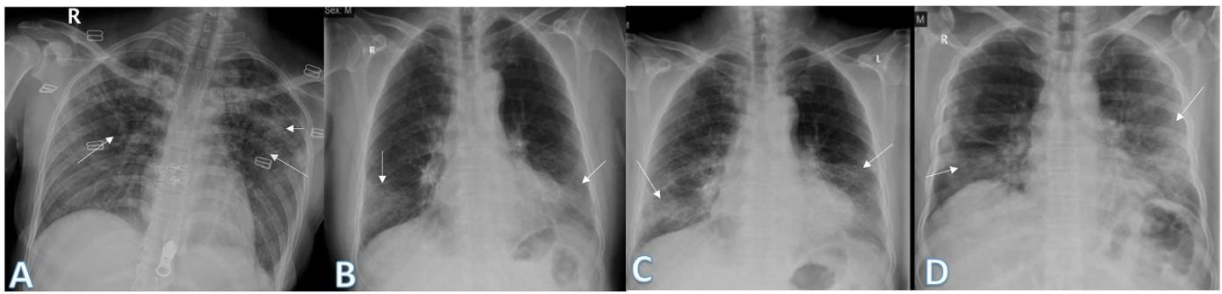

*Fig 5 [49]: Four different patients with positive COVID-19. (A) 29-year female. Initial chest X-ray showed bilateral upper zonal ground glass opacities (long arrows) with small left upper zonal air space consolidation opacities (short arrow). (B, C) 74-year male. Initial and 1st follow-up chest X-ray showed bilateral lower zonal ground glass opacities (arrows). (D) 54-year male. Initial chest X-ray show showed right lower zonal and left mid and lower zonal ground glass opacities (arrows).*

In our study, we experimented with some common Convolutional Neural Network models like Le Net-5, VGG-16, Res Net-50, Alex Net, and Inception-v3.

Although ResNet and Inception are much deeper neural networks, we decided to use VGG-16 for our work. The reason being, primary covid-19 indicators, 'ground-glass opacities' is a pixel-level feature that is present in the middle of the lungs, and extremely deep neural networks like ResNet tend to learn high-level features like bones, ECG wires, pacemakers, etc. which are not desired in this case.

We further removed the final two layers from our VGG-16 architecture for similar reasons. Finally, a 3 unit soft-max layer was added to the network. The network architecture and model summary are presented in Fig 6 and Table 3 respectively.

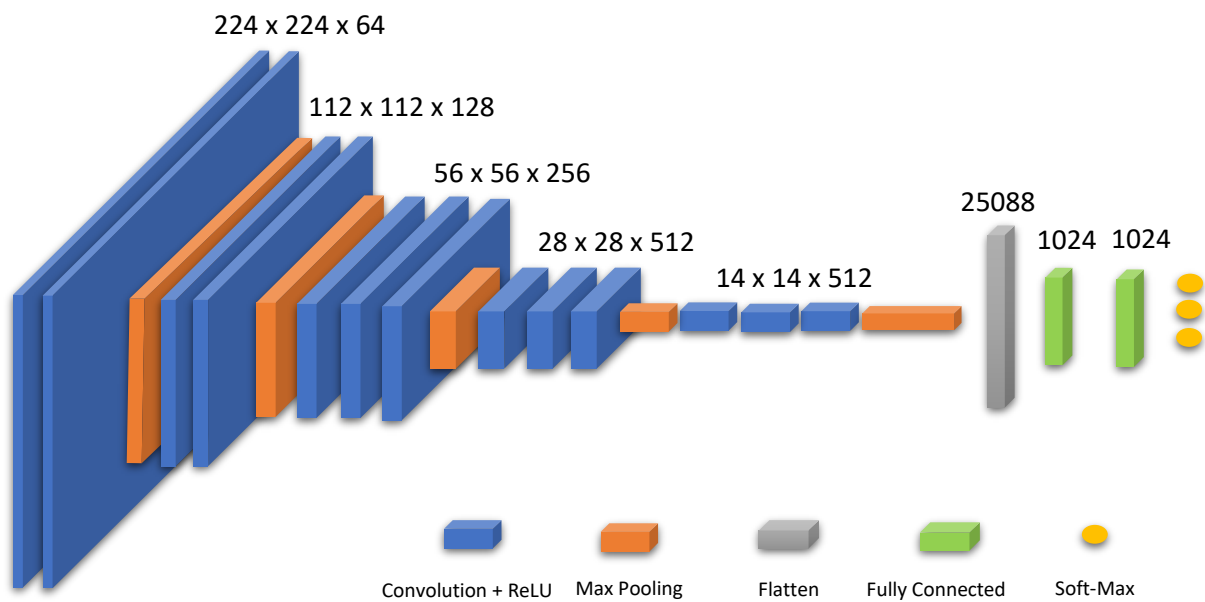

*Fig 6: Network Architecture for feature extraction*

*Table 3: Structure and parameters of proposed architecture.*
*(Conv – Convolution; MP –Max Pooling; FC – Fully Connected; R- Relu)*

| Layer No. | Layer Type | Feature Map | Size | Kernel Size | Stride | Activation |
|---|---|---|---|---|---|---|
| 0 | Input Image | 1 | 224 x 224 x 3 | - | - | - |
| 1 | Conv | 64 | 224 x 224 x 64 | 3 x 3 | 1 | R |
| 2 | Conv | 64 | 224 x 224 x 64 | 3 x 3 | 1 | R |
| 3 | MP | 64 | 112 x 112 x 64 | 3 x 3 | 2 | R |
| 4 | Conv | 128 | 112 x 112 x 128 | 3 x 3 | 1 | R |
| 5 | Conv | 128 | 112 x 112 x 128 | 3 x 3 | 1 | R |
| 6 | MP | 128 | 56 x 56 x 128 | 3 x 3 | 2 | R |
| 7 | Conv | 256 | 56 x 56 x 256 | 3 x 3 | 1 | R |
| 8 | Conv | 256 | 56 x 56 x 256 | 3 x 3 | 1 | R |
| 9 | Conv | 256 | 56 x 56 x 256 | 3 x 3 | 1 | R |
| 10 | MP | 256 | 28 x 28 x 256 | 3 x 3 | 2 | R |
| 11 | Conv | 512 | 28 x 28 x 512 | 3 x 3 | 1 | R |
| 12 | Conv | 512 | 28 x 28 x 512 | 3 x 3 | 1 | R |
| 13 | Conv | 512 | 28 x 28 x 512 | 3 x 3 | 1 | R |
| 14 | MP | 512 | 14 x 14 x 512 | 3 x 3 | 2 | R |
| 15 | Conv | 512 | 14 x 14 x 512 | 3 x 3 | 1 | R |
| 16 | Conv | 512 | 14 x 14 x 512 | 3 x 3 | 1 | R |
| 17 | Conv | 512 | 14 x 14 x 512 | 3 x 3 | 1 | R |
| 18 | MP | 512 | 7 x 7 x 512 | 3 x 3 | 2 | R |
| 19 | FC | - | 25088 | - | - | R |
| 20 | FC | - | 1024 | - | - | R |
| 21 | FC | - | 1024 | - | - | R |
| 22 | FC | - | 3 | - | - | SoftMax |

The model was initialized using the weights of VGG-16 trained on the ImageNet dataset [50]. Finally, the model was trained from scratch on 60% of data and validated on 10 % of data. We kept 30% of the test data separate to prevent overfitting later at the classification stage. The details of various parameters used during the training are mentioned in Table 4.

*Table 4: Hyperparameter Settings*

| Parameter | Value |
| --- | --- |
| Batch Size | 32 |
| Class Weight | COVID-19: 10<br>Normal: 8<br>Pneumonia: 9 |
| Epochs | 30 |
| Learning Rate | 0.0001 |
| Optimizer | Stochastic Gradient Descent |
| Momentum | 0.9 |

After training the model, the softmax layer was removed and the rest of the model was used as an encoder. All the images in the dataset were then encoded into 1024-dimensional feature vectors using the values of the penultimate layer of the trained network. We then used Synthetic Minority Over Sampling Technique (SMOTE) [51,52] to produce a balanced dataset consisting of 1000 feature vectors (1024–dimensional) of each class. SMOTE is a widely used data augmentation technique that generates synthetic instances by selecting points that are close in the feature space of the minority class. To study the separability of data at this stage, we plotted the projections of generated feature vectors using t-distributed stochastic neighbor embedding (t-SNE) [53] and Uniform Manifold Approximation and Projection (UMAP) [54] in Fig 7 (a) and (b) respectively. UMAP and t-SNE are popular dimensionality reduction techniques that are used to visualize higher-dimensional features. We can observe from Fig-7 that the COVID-19 class features are significantly separable from normal and pneumonia class features. However, there is a significant overlap between Normal and Pneumonia features. This can be attributed to the corresponding class weight. The model focuses more on differentiating COVID-19 samples from the other two classes than on differentiating all three classes.

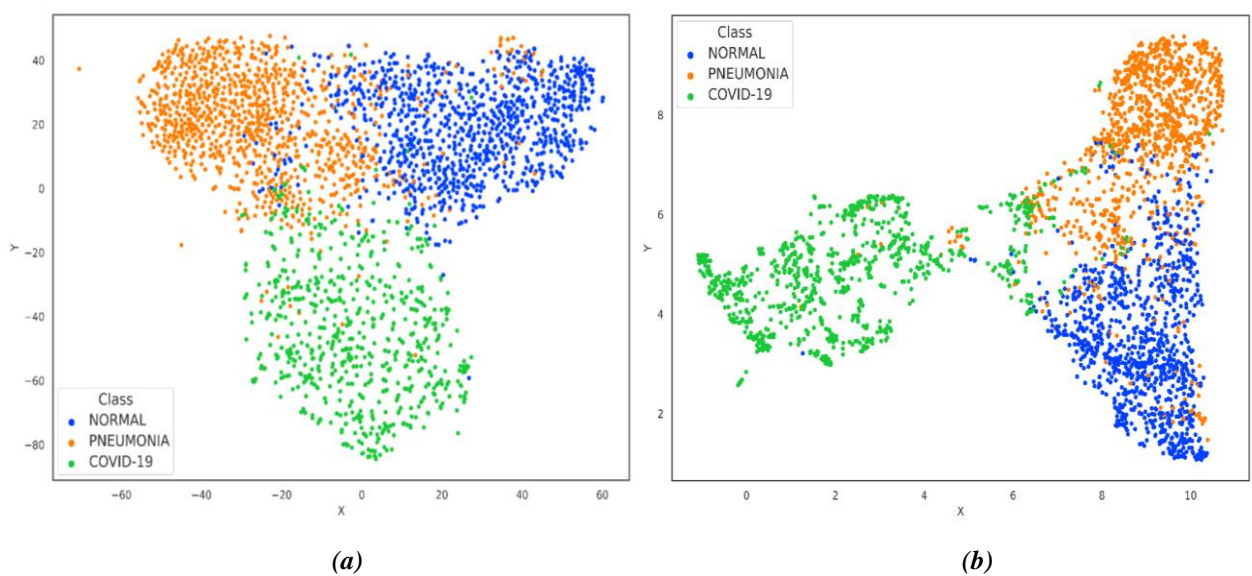

*(a)*         *(b)*

*Fig 7 (a) t-SNE projection (b) UMAP projection*

*2.4 Classification and Evaluation*

Finally, we used the Support Vector Machine Classifier with RBF kernel for multi-class classification of balanced data obtained in the last step. As a cross-validation method, k-fold was chosen with k=10. The confusion matrix for each fold was recorded.

Quantitative performance of the model was then evaluated using various performance metrics such as- Sensitivity (Se), Specificity (Sp), Accuracy (ACC), F-Score (F-Score), and Precision (Pre). These metrics were calculated using True Positive (TP), True Negative (TN), False Positive (FP), and False Negative (FN) parameters generated from each class of the confusion metrics.

$$Se = \frac{TP}{TP+FN} \tag{1}$$

$$Sp = \frac{TN}{FP+TN} \tag{2}$$

$$Acc = \frac{TP+TN}{TP+FN+TN+FP} \tag{3}$$

$$Pre = \frac{TP}{TP+FP} \tag{4}$$

$$F - Score = \frac{2*TP}{2*TP+FP+FN} \tag{5}$$

Sensitivity (Se) is the probability of a positive test result given that the patient has a condition. Specificity (Sp) is the probability of a negative test result given that the patient does not have the condition. Accuracy (ACC) is a measure for how many correct predictions the model made for the complete test dataset. F1-Score (F-Score) is a measure of model's accuracy based on sensitivity and precision. It is essentially the harmonic mean of the sensitivity and precision of the model. Precision (Pre) is a measure for the correctness of positive predictions made by the model. True Positive (TP) is a test result that correctly indicates the presence of a condition or characteristic. True Negative (TN) is a test result that correctly indicates the absence of a condition or characteristic. False Positive (FP) is a test result that wrongly indicates the presence of a condition or characteristic. False Negative (FN) is a test result that correctly indicates the absence of a condition or characteristic.

## 3. Results

In this section, we discuss the results of the proposed methodology. Table 5 shows the performance of various models we tried out in our work. Clearly, VGG outperforms other owing to it shallow network.

*Table 5: Support, Se. Sp., Pre, F1-score, and Acc. values of the proposed model for COVID-19 class*

| Model | Training Accuracy | Validation Accuracy |
|---|---|---|
| LeNet | 78.11 | 74.75 |
| AlexNet | 73.43 | 70.04 |
| ResNet-50 | 81.21 | 75.71 |
| VGG-16 | 92.86 | 87.34 |
| Inception-V3 | 90.11 | 84.29 |
| Xception | 70.30 | 67.76 |

After finalizing VGG, we further performed remaining calculations. For each fold, we have calculated the multi-class classification performance of the proposed framework. Further, we have also evaluated the average for various performance metrics both fold wise and class wise. The confusion matrix obtained for each fold during the 10-fold cross-validation is shown in Fig. 8 (b) to (k). The overlapped confusion matrix shown in Fig. 8 (a) is calculated using the sum of confusion matrices for all folds.

The values of various evaluation metrics of each fold for COVID-19, Normal, and Pneumonia class are tabulated in Tables 6, 7, and 8 respectively. The support-weighted average of performance metrics for each fold is mentioned in Table 9. Finally, the classification performance metrics values for the overlapped confusion matrix are shown in table 10. Clearly, from table 6, we can note that sensitivity, specificity, accuracy, precision, and F1-Score achieved by the proposed model for classification of COVID-19 are remarkably high. The average sensitivity, specificity, and precision are 98.71%, 99.01%, and 98% while the average accuracy and F1-Score are 98.9% and 98.34% respectively.

## Overlap

|  | COVID-19 | Normal | Pneumonia |
|---|---|---|---|
| **COVID-19** | 98.7% 987/1000 | 0.3% 3/1000 | 1.0% 10/1000 |
| **Normal** | 0.4% 4/1000 | 93.7% 937/1000 | 5.9% 59/1000 |
| **Pneumonia** | 1.6% 16/1000 | 7.2% 72/1000 | 91.2% 912/1000 |

Actual Label / Predicted Label

(a)

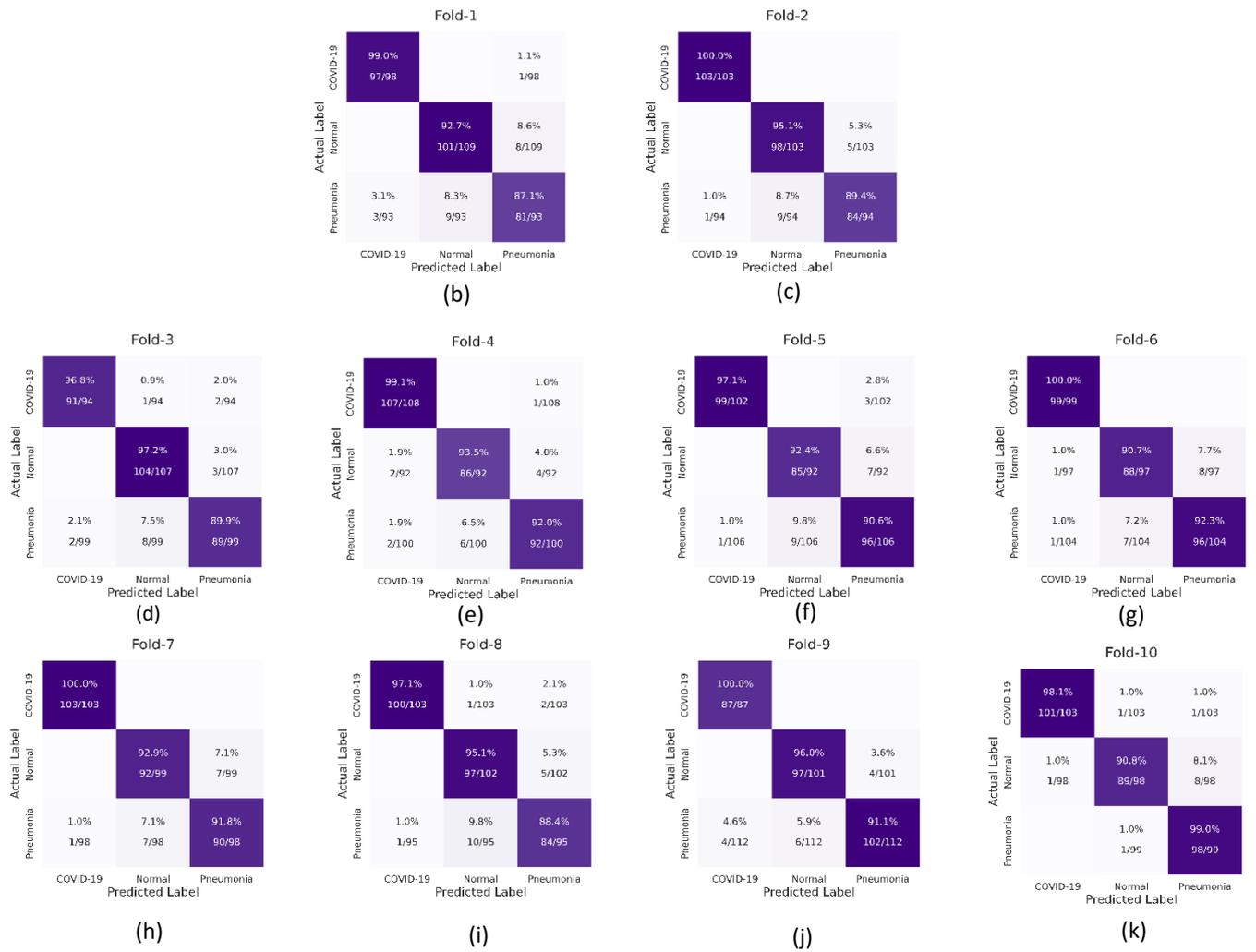

*Fig 8 (a) Overlapped Confusion Matrix; (b)-(k) 10-fold Confusion Matrix*

*Table 6: Support, Se. Sp., Pre, F1-score, and Acc. values of the proposed model for COVID-19 class*

| Fold | Support | Sensitivity | Specificity | Precision | Accuracy | F1-Score |
|---|---|---|---|---|---|---|
| 1 | 98 | 98.98 | 98.51 | 97 | 98.67 | 97.98 |
| 2 | 103 | 100 | 99.49 | 99.04 | 99.67 | 99.52 |
| 3 | 94 | 96.81 | 99.03 | 97.85 | 98.33 | 97.33 |
| 4 | 108 | 99.07 | 97.92 | 96.4 | 98.33 | 97.72 |
| 5 | 102 | 97.06 | 99.49 | 99 | 98.67 | 98.02 |
| 6 | 99 | 100 | 99 | 98.02 | 99.33 | 99 |
| 7 | 103 | 100 | 99.49 | 99.04 | 99.67 | 99.52 |
| 8 | 103 | 97.09 | 99.49 | 99.01 | 98.67 | 98.04 |
| 9 | 87 | 100 | 98.12 | 95.6 | 98.67 | 97.75 |
| 10 | 103 | 98.06 | 99.49 | 99.02 | 99 | 98.54 |
| Average | 100 | 98.71 | 99.01 | 98 | 98.9 | 98.34 |

Table 7: Support, Se. Sp., Pre, F1-score, and Acc. values of the proposed model for Normal class

| Fold | Support | Sensitivity | Specificity | Precision | Accuracy | F1-Score |
|---|---|---|---|---|---|---|
| 1 | 93 | 87.1 | 95.65 | 90 | 93 | 88.52 |
| 2 | 94 | 89.36 | 97.57 | 94.38 | 95 | 91.8 |
| 3 | 99 | 89.9 | 97.51 | 94.68 | 95 | 92.23 |
| 4 | 100 | 92 | 97.5 | 94.85 | 95.67 | 93.4 |
| 5 | 106 | 90.57 | 94.85 | 90.57 | 93.33 | 90.57 |
| 6 | 104 | 92.31 | 95.92 | 92.31 | 94.67 | 92.31 |
| 7 | 98 | 91.84 | 96.53 | 92.78 | 95 | 92.31 |
| 8 | 95 | 88.42 | 96.59 | 92.31 | 94 | 90.32 |
| 9 | 112 | 91.07 | 97.87 | 96.23 | 95.33 | 93.58 |
| 10 | 99 | 98.99 | 95.52 | 91.59 | 96.67 | 95.15 |
| Average | 100 | 91.16 | 96.55 | 92.97 | 94.77 | 92.02 |

Table 8: Support, Se. Sp., Pre, F1-score, and Acc. values of the proposed model for Pneumonia class

| Fold | Support | Sensitivity | Specificity | Precision | Accuracy | F1-Score |
|---|---|---|---|---|---|---|
| 1 | 109 | 92.66 | 95.29 | 91.82 | 94.33 | 92.24 |
| 2 | 103 | 95.15 | 95.43 | 91.59 | 95.33 | 93.33 |
| 3 | 107 | 97.2 | 95.34 | 92.04 | 96 | 94.55 |
| 4 | 92 | 93.48 | 97.12 | 93.48 | 96 | 93.48 |
| 5 | 92 | 92.39 | 95.67 | 90.43 | 94.67 | 91.4 |
| 6 | 97 | 90.72 | 96.55 | 92.63 | 94.67 | 91.67 |
| 7 | 99 | 92.93 | 96.52 | 92.93 | 95.33 | 92.93 |
| 8 | 102 | 95.1 | 94.44 | 89.81 | 94.67 | 92.38 |
| 9 | 101 | 96.04 | 96.98 | 94.17 | 96.67 | 95.1 |
| 10 | 98 | 90.82 | 99.01 | 97.8 | 96.33 | 94.18 |
| Average | 100 | 93.65 | 96.24 | 92.67 | 95.4 | 93.12 |

Table 9: Support, Se. Sp., Pre, F1-score, and Acc. values of the proposed model for each fold

| Fold | Support | Sensitivity | Specificity | Precision | Accuracy | F1-Score |
|---|---|---|---|---|---|---|
| 1 | 300 | 93 | 96.45 | 92.95 | 95.34 | 92.96 |
| 2 | 300 | 95 | 97.5 | 95.02 | 96.72 | 94.98 |
| 3 | 300 | 94.67 | 97.21 | 94.73 | 96.4 | 94.65 |
| 4 | 300 | 95 | 97.53 | 94.98 | 96.73 | 94.98 |
| 5 | 300 | 93.33 | 96.68 | 93.39 | 95.56 | 93.36 |
| 6 | 300 | 94.33 | 97.14 | 94.3 | 96.21 | 94.31 |
| 7 | 300 | 95 | 97.54 | 94.98 | 96.71 | 94.99 |
| 8 | 300 | 93.67 | 96.86 | 93.76 | 95.83 | 93.67 |
| 9 | 300 | 95.33 | 97.65 | 95.36 | 96.75 | 95.3 |
| 10 | 300 | 96 | 98.02 | 96.17 | 97.36 | 95.99 |
| Average | 300 | 94.53 | 97.26 | 94.56 | 96.36 | 94.52 |

*Table 10: Support, Se. Sp., Pre, F1-score, and Acc. values of the proposed model for overlapped confusion matrix*

| Class | Support | Sensitivity | Specificity | Precision | Accuracy | F1-Score |
|---|---|---|---|---|---|---|
| Normal | 1000 | 93.7 | 96.25 | 92.59 | 95.4 | 93.14 |
| COVID-19 | 1000 | 98.7 | 99 | 98.02 | 98.9 | 98.36 |
| Pneumonia | 1000 | 91.2 | 96.55 | 92.97 | 94.77 | 92.07 |
| Average | 1000 | 94.54 | 97.27 | 94.53 | 96.36 | 94.57 |

The overall performance of the model can be estimated using Tables 8 and 9.

## 4. Discussion

COVID-19 pandemic is getting worse day by day and there is a need for quick testing and diagnosis of patients. In this work, we propose a novel framework for the reliable detection of COVID-19 using chest radiographs. During the course of this study, we have tried to identify major drawbacks and flaws in the existing studies related to COVID-19 detection and classification using chest X-ray images. In this section, we shall discuss some of the existing studies alongside their advantages and limitations.

For this study, we sampled around 2470 images (1000 Pneumonia, 470 COVID-19, and 1000 Normal) from the COVIDx dataset. Most of the studies that we shall discuss in this section have used data from this dataset or the sources of this dataset.

COVID-Net designed by L.Wang et al [28] achieved a 93.3% accuracy using 358 COVID-19 images and 13617 non-COVID-19 images. However, in their study, they used horizontal flipping as one of the methods for data augmentation. We would like to point out the use of horizontal or vertical flipping is logically flawed in this context. Since, during the testing & inference process, X-rays will never be flipped. In their study, they used GSInquire to validate if the model is learning features outside of the lung. F. Ucar and D. Korkmaz [31] employed a fine-tuned DeepSqueezeNet with Bayesian Optimization on 76 COVID-19 infected lungs images and 5873 for non-COVID-19 ones. Their accuracy came around 98.26%. To handle class imbalance, they generated around 1500 images from the initial 76 images using various image augmentation methods. This may have led to high over-fitting. Considering the lack of data, J. Zhang et al. [26] assumed COVID-19 to be an anomaly detection problem and they diagnosed covid-19 by using a pre-trained ResNet-18 for feature extraction along with an anomaly detection head. They achieved an average 83.32% accuracy rate with 100 COVID-19 positive and 1431 non-COVID images. Their model can still learn features outside the lungs.

*Table 11: Comparison of the proposed model with other deep learning methods*

| Reference | COVID-19 Images | Non-COVID-19 Images | Technique Used | Accuracy |
|---|---|---|---|---|
| *L. Wang and Wong [28]* | 358 | 13617 | Tailor-made COVID-Net model | 93.3 |
| *F. Ucar et al. [31]* | 76 | 5873 | Fine-tuned DeepSqueezeNet with Bayesian optimization | 98.3 |
| *J. Zhang et al. [26]* | 100 | 1431 | ResNet-18 with anomaly detection head | 83.3 |
| *R. Kumar et al. [27]* | 62 | 8526 | ResNet-152 for feature extraction, SMOTE for oversampling, XGBoost for classification | 97.7 |
| *R. M. Pereira et al. [33]* | 90 | 1054 | Hierarchical Classification | 95.2 |
| *Chowdhury et al. [39]* | 190 | 2686 | SqueezeNet | 98.3 |
| *Apostolopoulos et al. [42]* | 224 | 1204 | VGG-19 | 93.5 |
| *Sitaula et al. [55]* | 320 | 320 | Bag of Visual Words (BoVW) | 87% |
| **Proposed Method** | **470** | **2000** | **Lung Segmentation, VGG-16 for feature extraction, SMOTE oversampling, Support Vector Classifier** | **96.4** |

R. Kumar et al. [27] used ResNet-152 for feature extraction and SMOTE for oversampling the data. The obtained 97.7% accuracy. It is worth noting that they have generated 5000 COVID-19 class samples from 62 samples during oversampling and then used XGBoost for classification. R. M. Pereira et al. [33] experimented with a variety of feature extraction techniques and formulated the problem as hierarchal classification. They obtained an accuracy of 98.66 % using 90 COVID-19 images. They also performed an edge cut to avoid detecting features outside the lung. Loey et al. [34] used GAN (generative adversarial network) to generate more samples of COVID-19 images. Chowdhury et al. [39] curated a dataset with 190 COVID-19 images from different sources and obtained a 98.3 % accuracy using SqueezeNet. Apostolopoulos et al. [42] have used 224 COVID-19 images and VGG-19. They achieved an accuracy of 93.48 %. The various studies differ in the following aspects – Size of the dataset used, handling the lack of data & class imbalance, and the method used for feature extraction.

It can also be observed that major drawbacks of related studies include-

- *Lack of training data*- Some studies have used as few as 50 images.
- *Extreme oversampling*- A few studies have generated as much as 100 times more samples from existing samples.
- *Impractical data augmentation*- A few studies have used horizontal or vertical flipping for data augmentation.
- *Irrelevant feature extraction*- In a lot of studies, the model may learn features outside of the lungs to classify the X-ray image as COVID-19.
- *Distinctive Data Source*- Some studies have used a dataset that is inherently distinctive as described in previous section.

In our study, we tried to address all these challenges. The main advantages of the proposed model are-

- *Comparatively more data* – In this work, we have made use of 470 COVID-19 images.
- *Avoiding impractical data augmentation*- We have avoided the use of methods such as horizontal flipping, random crop, rotation, translation, etc.
- *Limited Oversampling* – We generated 1000 samples from 470 samples.
- *The lung segmentation* technique is used so that no feature will be detected outside the lungs.

For COVID-19, the proposed framework obtained an accuracy of 98.9%, 94.7% for pneumonia class, and 95.4% for the normal class. It is worth noting that the average accuracy for normal and pneumonia class (Table 7 and 8) is relatively lower than that of COVID-19. This can be attributed to the higher-class weight of COVID-19 during the feature extraction phase. This resulted in high penalization for the misclassification of COVID-19 classes. In Table 11, we compare several of earlier models with our proposed framework. It is evident from Table 10 that the performance of the proposed method is at par with the existing methods. Our method has managed to achieve remarkably high accuracy with relatively more data than other studies. We now present some limitation and prospects of our work-

- Although the current study has used a relatively larger dataset, COVID-19 images are still less in number.
- The reliability and explain-ability of the model can be further improved using infection segmentation instead of lung segmentation.
- We currently use a modified VGG-16 for feature extraction. The accuracy can be improved by designing a custom neural network for the same.
- The accuracy can be further improved by incorporating an ensemble of various feature extraction techniques.
- The proposed model needs to be validated on larger dataset.

## 5. Conclusion

As COVID-19 increase rapidly, quick and reliable diagnosis of the disease is needed. Current diagnosis techniques like RT-PCR are time-consuming and have low sensitivity. Recent studies show that COVID-19 can be reliably diagnosed using a chest X-Ray. Therefore, we propose a novel deep learning approach for classification of COVID-19 using chest X-ray images. We tried to address the major limitations and drawbacks of previous work done in this direction. Our novel approach combining well-known pre-processing techniques like Histogram Equalization and U-net lungs segmentation technique, lead us to an outstanding rate of recognition of 98% for 470 COVID-19 X-ray images over a total dataset of 2470 X-ray images. Future work could include further segmentation of infected lungs region and validation over a larger dataset.

## Conflict of Interest / Competing Interests

The authors declare that they have no conflict of interest. Further, the authors have no relevant financial or non-financial interests to disclose. No funds, grants, or other support was received.

## Acknowledgement

The following work is an revised version of the preprint [56].

[47] Ronneberger, Olaf, Philipp Fischer, and Thomas Brox. "U-net: Convolutional networks for biomedical image segmentation." *International Conference on Medical image computing and computer-assisted intervention*. Springer, Cham, 2015.

[48] W. Kong and P. P. Agarwal, "Chest Imaging Appearance of COVID-19 Infection," Radiology: Cardiothoracic Imaging, vol. 2, no. 1, p. e200028, Feb. 2020, DOI: 10.1148/ryct.2020200028.

[49] R. Yasin and W. Gouda, "Chest X-ray findings monitoring COVID-19 disease course and severity," Egypt J Radiol Nucl Med, vol. 51, no. 1, Sep. 2020.

[50] A. Krizhevsky, I. Sutskever, G.E. Hinton, ImageNet classification with deep convolutional neural networks, in Advances in Neural Information Processing Systems, 2012, pp. 1097–1105.

[51] Kovács, G. An empirical comparison, and evaluation of minority oversampling techniques on a large number of imbalanced datasets. Applied Soft Computing 83, 105662 (2019).

[52] Fernández, A., Garcia, S., Herrera, F. & Chawla, N. V. Smote for learning from imbalanced data: progress and challenges, marking the 15-year anniversary. Journal of artificial intelligence research 61, 863–905 (2018).

[53] L. van der Maaten, G. Hinton, Visualizing data using t-SNE, J. Mach. Learn. Res. 9 (2008) 2579–2605.

[54] McInnes et al., (2018). UMAP: Uniform Manifold Approximation and Projection. Journal of Open Source Software, 3(29), 861, https://doi.org/10.21105/joss.00861

[55] Sitaula, C., & Aryal, S. (2021). New bag of deep visual words based features to classify chest x-ray images for COVID-19 diagnosis. Health Information Science and Systems, 9(1), 1-12.

[56] Bhadouria, Hitendra Singh, et al. "Classification of COVID-19 on chest X-Ray images using Deep Learning model with Histogram Equalization and Lungs Segmentation." *arXiv preprint arXiv:2112.02478* (2021).

## Author Biography

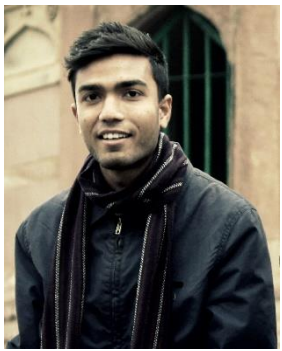

**Aman Swaraj**
PhD scholar, Computer Science & Engineering
Indian Institute of Technology, Roorkee

**Research Interest:** Machine learning, Deep Learning in Image Recognition, and Time Series Forecasting.

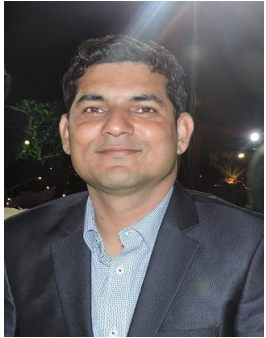

**Dr. Karan Verma**
Assistant Professor CSE, NIT Delhi
PhD (2015) PETRONAS, Malaysia.
**Research Interest:** ML, Deep Learning, Information security, Wireless Sensor Networks and VANET.